\documentclass[pra, nofootinbib, usenatbib, showkeys, showpacs, notitlepage]{revtex4-1}
\usepackage[utf8]{inputenc} 
\usepackage[english]{babel}
\usepackage{mathrsfs}
\usepackage{fullpage}
\usepackage{graphicx}
\usepackage{fancyvrb}
\usepackage{units}
\usepackage{xcolor}
\usepackage{lmodern}
\usepackage{amsmath,amsfonts}
\usepackage{array}
\usepackage{indentfirst}
\usepackage{perpage} 
\MakePerPage{footnote} 
\usepackage[babel=true]{csquotes} 
\usepackage{hyperref}

\usepackage{minitoc}

\usepackage{natbib}

\begin{document}

\title{Landau levels for discrete-time quantum walks in artificial magnetic fields}

\author{Pablo Arnault}
\email{pablo-arnault@hotmail.fr}

\author{Fabrice Debbasch}
\email{fabrice.debbasch@gmail.com}


\affiliation{LERMA, Observatoire de Paris, PSL Research University, CNRS,\\
Sorbonne Universit\'es, UPMC Univ. Paris 6, UMR 8112, F-75014, Paris France}


\begin{abstract}
A new family of $2D$ discrete-time quantum walks (DTQWs) is presented and shown to coincide, in the continuous limit, with the Dirac dynamics of a spin 1/2 fermion coupled to a constant and uniform magnetic field. Landau levels are constructed, not only in the continuous limit, but also for the DTQWs {\sl i.e.} for finite non-vanishing values of the time- and position-step, by a perturbative approach in the step. Numerical simulations support the above results and 
suggest that the magnetic interpretation is valid beyond the scope of the continuous limit. The possibility of quantum simulation of condensed-matter systems by DTQWs is also discussed.
\end{abstract}



\maketitle

\section{Introduction}

Quantum walks (QWs) were introduced in the litterature by Y. Aharonov \cite{ADZ93a} and D. A. Meyer \cite{Meyer96a}. They are models of coherent quantum transport in discrete space. This article focuses on Discrete-Time Quantum Walks (DTQWs), which are quantum formal analogues to Classical Random Walks (CRWs). A typical $1D$ quantum walker carries a two-state internal degree of freedom or spin. At each time-step of a DTQW,  the spin of the walker is rotated deterministically through a so-called coin operator and the walker then undergoes a spin-dependent position shift to the left or to the right. This evolution naturally entangles the spin and the position of the walker. The basic construction just outlined has been generalized to higher spatial dimensions and/or coin spaces of dimension higher than two \cite{MBSS2002, Arrighi_higher_dim_2014}. Two standard reviews on (DT)QWs are \cite{Kempe_review, Venegas_review}.

DTQWs have been realized experimentally  with a wide range of physical objects and setups
\cite{Sanders03a, Perets08a, Schmitz09a, Karski09a, Zahring10a, Schreiber10a,Sansoni2012}.
They are important in a large variety of contexts, ranging from fundamental quantum physics \cite{Perets08a, var96a} to quantum algorithmics \cite{Amb07a, MNRS07a,Childs2013,book_Barnett,book_Manouchehri, book_Sansoni}, solid state physics \cite{Aslangul05a, Bose03a, Burg06a, Bose07a} and biophysics \cite{Collini10a, Engel07a}. 

Contrary to CRWs, DTQWs are actually deterministic. Indeed, while CRWs describe diffusion,  DTQWs describe wave propagation at finite velocity. DTQWs are therefore linked with Lorentz geometry \cite{AFF14a} and in particular with the Dirac equation\footnote{Indeed, Feynmann \cite{FeynHibbs65a} developped in 1965 a QW-like path discretization for Dirac fermions.}. 

It has been shown recently \cite{DMD12a, DMD13b, DMD14} that several families of $1D$ DTQWs could be interpreted as discrete models of spin 1/2 Dirac fermions coupled to electric and relativistic gravitational fields. 
The aim of this article is to extend these results to magnetic fields. We thus introduce a new family of $2D$ DTQWs and show that the formal continuous limit of this family coincides with the Dirac dynamics of a spin 1/2 fermion coupled to a constant and homogeneous magnetic field orthogonal to the $2D$ space in which the walk propagates. 

The eigenvalues of the Hamiltonian generating  
this continuous dynamics are well-known and are called the relativistic Landau levels. We show by a perturbative computation that Landau levels also exist for the discrete dynamics of the DTQWs and discuss how the corresponding eigenstates and probability densities are influenced by the finite non-vanishing time- and position-step. All results are finally supported by numerical simulations. 

The results presented in this article strongly support that DTQWs can be used for quantum simulation of condensed-matter systems and spintronics. Indeed, much of the recent research on condensed-matter physics involves the $2D$ interaction of electrons with magnetic fields. Landau levels play a prominent role in this field and are for example one of the key ingredients of the quantum Hall effect \cite{Goerbig_review}, which has current important applications in metrology \cite{JJ01}, and could also be used in quantum computation \cite{DFN05, AG02, PVK98}. Thus, the existence of Landau levels for the DTQWs make DTQWs natural promising candidates for quantum simulation of many interesting phenomena, including the quantum Hall effect. 


\section{Fundamentals} \label{section_2DQW}


Consider the DTQW defined by the following set of discrete evolution equations:
\begin{eqnarray}
\left(\begin{array}{c}
\psi^L_{j+1/2, p, q} \\
\psi^R_{j+ 1/2, p, q}
\end{array}
\right)
& = & \mathbf{U}(\alpha_{p}(\nu, B), \theta^+(\nu, m))
\left(
\begin{array}{c}
\psi^L_{j, p+1, q} \\
\psi^R
_{j, p-1, q}
\end{array}
\right)
\nonumber \\
\left(\begin{array}{c}
\psi^L_{j+1, p, q} \\
\psi^R_{j+ 1, p, q}
\end{array}
\right)
& = & \mathbf{V}(\alpha_{p}(\nu, B), \theta^-(\nu, m))
\left(
\begin{array}{c}
\psi^L_{j+1/2, p, q+1} \\
\psi^R_{j+1/2, p, q-1}
\end{array}
\right),
\label{eq:fund}
\end{eqnarray}
where
\begin{equation} \label{best_form_bis}
\mathbf{U}(\alpha, \theta) =
  \left[
  \begin{array}{c c}
  e^{i\alpha} \cos \theta &
  i e^{i\alpha} \sin \theta \\
  i e^{-i\alpha} \sin \theta &
  e^{-i\alpha} \cos \theta
  \end{array}
  \right], \ \ \ \ \ \ \ \  
\mathbf{V}(\alpha, \theta) =
  \left[
  \begin{array}{c c}
  e^{i\alpha} \cos \theta &
  i e^{-i\alpha} \sin \theta \\
  i e^{i\alpha} \sin \theta &
  e^{-i\alpha} \cos \theta
  \end{array}
  \right] ,  
\end{equation} 
\begin{equation}
\alpha_{p} (\nu, B) = \nu^2 \, \frac{B p}{2},
\end{equation}
\begin{equation}
\theta^{\pm} (\nu, m) = \pm \frac{\pi}{4} - \nu \, \frac{m}{2} \ ,
\end{equation}
and $\nu$, $B$, $m$ are arbitrary parameters. The indices $j$ and $p,q$ are integers wich label discrete time and positions in $2D$ discrete space, respectively. The wave function $\Psi$ lives in a two-dimensional Hilbert space and is thus represented by a `spinor' $\Psi$ with components $(\psi^L, \psi^R)$ in a certain $(j,p,q)$-independent basis $(b_L,b_R)$ of this Hilbert space, {\sl i.e.} $\Psi = \psi^L b_L + \psi^R b_R$ .

To make notations less cumbersome, we write $\mathbf{U}^+(\nu) = \mathbf{U}(\alpha_{p}(\nu, B), \theta^+(\nu, m))$ and $\mathbf{V}^-(\nu) = \mathbf{U}(\alpha_{p}(\nu, B), \\ \theta^-(\nu, m))$. Of the three angles entering the definitions of $\mathbf{U}^+(\nu)$ and $\mathbf{V}^-(\nu)$, only one depends on the point at which the transport is computed, and only through the first space-coordinate $p$. The parameters $B$ and $m$ will eventually be interpreted as a magnetic field and as a  mass. The interpretation of the parameter $\nu$ will be discussed below.

The evolution from `time' $j$ to time $j+1$ proceeds in two steps. In the first step, the DTQW is transported by operator $\mathbf{U}^+(\nu)$ from time $j$ to time $j + 1/2$ along the first space direction ($p$-direction). In the second step, the DTQW is transported by operator $\mathbf{V}^-(\nu)$ from time $j+1/2$ to time $j + 1$ along the other space direction ($q$-direction). Note however that the values of $\Psi$ at times $j + 1/2, j \in \mathbb N$ are not taken into account in the DTQW {\sl i.e.} the DTQW is the succession of the values taken by $\Psi$ at all integer times $j$.
The values of $\Psi$ at half-integer times are thus merely a tool to advance the DTQW in `time'. Equations relating directly $\Psi_{j+1}$ to $\Psi_j$ can be deduced from (\ref{eq:fund}) and read:
\begin{align}
\label{eq:fund2}
\psi^L_{j+1,p,q} & = e^{2i\alpha_p} \cos \theta^- \big[ \cos \theta^+ \psi^L_{j,p+1,q+1} 
+ i \sin \theta^+ \psi^R_{j,p-1,q+1} \big] \nonumber \\
 & \ \ \ + ie^{-2i\alpha_p} \sin \theta^- \big[i \sin \theta^+ \psi^L_{j,p+1,q-1} 
+  \cos \theta^+ \psi^R_{j,p-1,q-1} \big] \ ,  \nonumber \\
 \\
\psi^R_{j+1,p,q} & = ie^{2i\alpha_p} \sin \theta^- \big[ \cos \theta^+ \psi^L_{j,p+1,q+1} 
+ i \sin \theta^+ \psi^R_{j,p-1,q+1} \big] \nonumber \\
 & \ \ \ + e^{-2i\alpha_p} \cos \theta^- \big[ i \sin \theta^+ \psi^L_{j,p+1,q-1} 
+  \cos \theta^+ \psi^R_{j,p-1,q-1} \big]  \ . \nonumber
\end{align}

\section{Continuous limit} \label{section:continuous_limit}

To determine the formal continuous limit of this DTQW, we proceed as in \cite{DMD14, DMD12a, DMD13b, DMR97a}. We first introduce (dimensionless) time and space steps $\Delta T$, $\Delta X$, $\Delta Y$ and interpret $\Psi_{j, p, q}$ and $\alpha_p$ as the value $\Psi(M)$ and $\alpha(M)$ taken by certain functions $\Psi$ and $\alpha$ at point $M$ with (dimensionless) space-time coordinates $T_j = j\, \Delta T$, $X_p = p\,  \Delta X$, $Y_q = q\,  \Delta Y$. Since DTQWs are essentially discrete waves, it makes sense to choose $\Delta T = \Delta X = \Delta Y$ (cubic lattice in space-time) \cite{DMD13b,DMD14} and denote by $\epsilon$ the common value of all three space-time steps. Equations (\ref{eq:fund2}) then transcribe into:
\begin{align}
\label{eq:fund3}
\psi^L(T_j+\epsilon,X_p,Y_q) & = e^{2i\alpha(X_p)} \cos \theta^- \big[ \cos \theta^+ \psi^L(T_j,X_p+\epsilon,Y_q+\epsilon) 
+ i \sin \theta^+ \psi^R(T_j,X_p-\epsilon,Y_q+\epsilon)  \big] \nonumber \\
& \ \ \ + ie^{-2i\alpha(X_p)} \sin \theta^- \big[ i \sin \theta^+ \psi^L(T_j,X_p+\epsilon,Y_q-\epsilon)  
+  \cos \theta^+ \psi^R(T_j,X_p-\epsilon,Y_q-\epsilon) \big]  \ ,\nonumber \\
 \\
\psi^R(T_j+\epsilon,X_p,Y_q) & = ie^{2i\alpha(X_p)} \sin \theta^- \big[ \cos \theta^+ \psi^L(T_j,X_p+\epsilon,Y_q+\epsilon)  
+ i \sin \theta^+ \psi^R(T_j,X_p-\epsilon,Y_q+\epsilon) \big] \nonumber \\
& \ \ \ + e^{-2i\alpha(X_p)} \cos \theta^- \big[ i \sin \theta^+ \psi^L(T_j,X_p+\epsilon,Y_q-\epsilon)
+  \cos \theta^+ \psi^R(T_j,X_p-\epsilon,Y_q-\epsilon) \big]  \ . \nonumber
\end{align}

We then let $\epsilon$ tend to zero. The differential equations fixing the dynamics of the formal continuous limit are determined by expanding 
(\ref{eq:fund3})
around $\epsilon = 0$ and balancing the zeroth- and first-order terms in $\epsilon$. 
The zeroth-order terms only balance if $\mathbf{V}^-(\nu)\mathbf{U}^+(\nu) \rightarrow 1$ when $\epsilon \rightarrow 0$. This condition implies that $\nu$ has to tend to zero with $\epsilon$. The most interesting scaling turns out to be $\nu = \epsilon$ and the first-order terms then deliver the following differential equations for $(\psi^L, \psi^R)$:
\begin{align}
\label{eq:continuous_limit_1}
(-iBX - \partial_Y - m)\psi^L + i (\partial_T + \partial_X) \psi^R & = 0 \nonumber \\
i (\partial_T - \partial_X)\psi^L + (iBX + \partial_Y - m)\psi^R & = 0 \ .
\end{align}

This system can be written in the simple so-called manifestly covariant form:
\begin{equation} \label{eq:Dirac}
(i\gamma^{\mu} D_{\mu} - m)\Psi = 0 \ ,  
\end{equation}
where the matrices
\begin{equation}
[{(\gamma^0)^a}_b]
= \sigma_1 = \left(
    \begin{array}{cc}
       0 & 1 \\
       1 & 0 
    \end{array}
  \right), \ \ \ \ \ \ \ \ [{(\gamma^1)^a}_b] = i \sigma_2 = \left(
    \begin{array}{cc}
       0 & 1 \\
       -1 & 0 
    \end{array}
  \right), \ \ \ \ \ \ \ \ [{(\gamma^2)^a}_b] = i \sigma_3 =  \left(
    \begin{array}{cc}
       i & 0 \\
       0 & -i 
    \end{array}
  \right) ,
\end{equation}
 $(a,b) \in \{L,R\}^2$, obey the $(2+1)$ dimensional flat space-time Clifford algebra relations $\gamma^{\mu} \gamma^{\nu} + \gamma^{\nu} \gamma^{\mu}= 2\eta^{\mu \nu}$ (with $\eta_{\mu \nu} = \mbox{diag}(1, -1, -1)$), and the covariant derivative $D$ is defined by $D_\mu = \partial_\mu - i A_\mu$, with 
\begin{equation} \label{eq:Landau_gauge}
A_0 = A_1 = 0 \ , \ \ \ \ A_2 = - B X \ . 
\end{equation} 
The vector potential $A$ generates a uniform magnetic field $B$ orthogonal to the $(X, Y)$ plane and equation (\ref{eq:Dirac}) is the Dirac equation describing the dynamics of a spin $1/2$ fermion of mass $m$ and charge $-1$ coupled to this magnetic field in $(2 + 1)$-dimensional space-time \cite{Goerbig_review, PPJI, Fuchs_HDR, Bernevig_book, Pascal_Simon_course}. To consider a particle of generic charge $q$, perform the substitution $B \rightarrow -qB$. 
The characteristic propagation speed of the Dirac equation (\ref{eq:Dirac}) is $1$ (remember the time and space variables $T$ and $X$ are dimensionless).
In practice, the Dirac equation is used to describe $(i)$ `actual' spin 1/2 particles at high energy~
\cite{Weinberg_QFT1}, in which case the characteristic speed is the speed of light $c$, and $(ii)$ low-energy excitations of graphene-like materials (also called Dirac materials), whose dispersion relation can by linearised at low energy around the so-called Dirac points \cite{Goerbig_review}, in which case the characteristic speed $v$ is the so-called Fermi velocity of the material (in the case of graphene, $v \simeq c/300$). We stress that, in condensed matter, the Dirac equation is thus only an effective description of long wavelength/small wave number excitations and that the Dirac equation certainly becomes invalid at wave numbers comparable to the size of the Brillouin zone.

\section{Landau levels} \label{section:Landau_levels}

\subsection{Hamiltonian of the DTQW}
\label{ssec:Ham}

The eigenvalues of the Hamiltonian generating the continuous limit dynamics (\ref{eq:Dirac}) are the well-known relativistic or Dirac Landau levels. We will now determine the eigenvalues and the eigenstates of the original DTQW for small but finite values of the time- and space-step $\epsilon$. 

To be definite, we suppose that space is infinite in both directions and introduce a discrete Fourier transform in the $q$- ({\sl i.e.} $Y$-) direction. We thus write, for all functions $f$ defined on space-time:
\begin{equation}
\hat{f} (T_j, X_p, K) = \sum_{q \in \mathbb Z} f(T_j, X_p, Y_q) \exp(-i K Y_q) \ ,
\end{equation}
and the inversion relation reads
\begin{equation}
f(T_j, X_p, Y_q) = \frac{1}{2 \pi} \ \int_{- \pi/\epsilon}^ {\pi/\epsilon} \hat{f} (T_j, X_p, K) \exp(i K Y_q) dK .
\end{equation}
We recall that $Y_q = q \epsilon$, $q \in \mathbb Z$ and the associated wave-vector $K$ is thus a continuous real variable which takes all values in the interval $[- \pi/\epsilon,  \pi/\epsilon]$. As $\epsilon$ tends to zero, the discrete transform goes into the standard continuous Fourier transform. 

Assuming analyticity and considering directly the scaling $\nu = \epsilon$, the exact discrete dynamics of the DTQW reads in Fourier space:

\begin{align}
\hat{\Psi}(T_j+\epsilon,X_p,K) & = 
{\mathbf W}(\epsilon, B, \theta^+ (\epsilon, m), \theta^- (\epsilon, m), X_p - 
\chi(K, B)) \left( \mathbf{D}(\epsilon)  \hat{\Psi} \right)_{(T_j,X_p,K)}
\nonumber 
\\
& =  \left( \mathbf{{Q}}(\epsilon, B, m)   \hat{\Psi} \right)_{(T_j,X_p,K)} \ ,
\label{eq:DTQWFourier}
\end{align}
where
\begin{equation}
\chi(K, B) = - \, \frac{K}{B},
\end{equation}
\begin{equation}
[{{\mathbf W}^a}_{ b} (\epsilon, B, \theta^+, \theta^-, X)] = 
\left[
\begin{array}{c c}
e^{i\epsilon B X} \cos \theta^- \cos \theta^+ 
          - e^{-i\epsilon B X} \sin \theta^- \sin \theta^+ &
i \left[ e^{i\epsilon B X} \cos \theta^- \sin \theta^+ 
          + e^{-i\epsilon B X} \sin \theta^- \cos \theta^+  \right] \\
i \left[ e^{i\epsilon B X}  \sin \theta^- \cos \theta^+
          + e^{-i\epsilon B X} \cos \theta^- \sin \theta^+  \right] &
- e^{i\epsilon B X} \sin \theta^- \sin \theta^+
          + e^{-i\epsilon B X} \cos \theta^- \cos \theta^+ 
\end{array}
\right],
\end{equation}
and 
\begin{equation}
[{\mathbf{D}^a}_{ b}({\epsilon})] =  
\left[
\begin{array}{c c}
 \exp(\epsilon \partial_X) &
0 \\
0 & \exp(- \epsilon \partial_X)
 
\end{array}
\right] ,
\end{equation}
\noindent
with $(a,b) \in \{L,R\}^2$. Equation (\ref{eq:DTQWFourier}) defines the operator $\mathbf{{Q}}(\epsilon, B, m)$.

If a Hamiltonian ${{\mathbf H}}(\epsilon, B, m)$
generates the dynamics (\ref{eq:DTQWFourier}), then 
\begin{eqnarray}
\hat{\Psi}(T_j+\epsilon,X_p,K)  = \left( \exp( - i \epsilon {{\mathbf H}}(\epsilon, B, m) )  \hat{\Psi} \right)_{(T_j,X_p,K)} .
\label{eq:defHFourier}
\end{eqnarray}

The existence of the formal continuous limit presented in the preceding section is traced by the fact that, for all values of $B$ and $m$, the operator ${\mathbf{Q}}(\epsilon, B, m)$ tends to unity as $\epsilon$ tends to zero. Thus, the logarithm of ${\mathbf{Q}} (\epsilon, B, m)$ is defined, at least for small enough values of $\epsilon$. A solution to (\ref{eq:defHFourier}) is therefore
\begin{equation}
{{\mathbf H}}(\epsilon, B, m)=  \frac{i}{\epsilon} \ln \mathbf{{Q}} (\epsilon, B, m) \ ,
\label{eq:defH}
\end{equation}
with
\begin{equation} \label{eq:log_Q}
\ln {\mathbf{Q}} =  \sum_{n=1}^{+\infty} \frac{(-1)^{n-1}}{n} ({\mathbf{Q}} - 1)^n = \sum_{n=1}^{+\infty} \sum_{k=0}^{n} \frac{(-1)^{k + 1}}{n} \binom{n}{k}  {{\mathbf{Q}}}^{k}.
\end{equation}

Let ${\mathbf{H}}(\epsilon, B, m) = \sum_{k=0}^{+\infty} \epsilon^k {\mathcal H}^{(k)} (B, m)$ be the expansion of ${{\mathbf H}}(\epsilon, B, m)$ in powers of $\epsilon$. The expression of some $\mathcal{H}^{(k)} (B, m)$ can be obtained by extracting all the terms of order $\epsilon^k$ in (\ref{eq:log_Q}). It turns out (see Appendix A) to be convenient to represent the Hamiltonian, not in the original basis $(b_L, b_R)$ chosen in spin state, but in the new basis $(b_-, b_+)$ defined by 
\begin{equation}
b_-  = \frac{1}{\sqrt{2}}(b_L + b_R) \ , 
\ \ \ \  \ \ \ \ \ \ \ \ \ \ \ \  \ 
b_+  = \frac{1}{\sqrt{2}}(-b_L + b_R) \ .
\end{equation}
In this new basis, we find
\begin{equation} \label{eq:Hamiltonian_form_TF_final}
[{(\mathcal{H}^{(0)})^u}_v (B, m)]_{(X,K)} = 
\left[
  \begin{array}{c c}
  m &
  -i\left(\partial_X + B( X - \chi(K, B))\right) \\
  i\left(-\partial_X + B(X - \chi(K, B))\right)&
  -m
  \end{array}
\right]  
\end{equation}
and
\begin{align}
\label{eq:hamiltonian_perturbation}
[({\mathcal{H}^{(1)})^u}_v (B, m)]_{(X,K)}  =
\left[
  \begin{array}{c c}
  iB[1/2 + ( X - \chi(K, B)) \partial_X] &
  - m \partial_X \\
  m \partial_X &
  -i B [1/2 + ( X - \chi(K, B))\partial_X]
  \end{array}
\right], 
\end{align}
\noindent
where $(u,v) \in \{-,+\}^2$.

The Landau levels of the DTQWs are, by definition, the eigenvalues of the Hamiltonian ${\bf H}(\epsilon, B, m)$. These levels are related to the so-called eigenvalues of the DTQW, which are by definition \cite{SK10, Shikano2011} the eigenvalues of the operator $\exp( - i \epsilon\,  {{\mathbf H}}(\epsilon, B, m) )$. Thus, if $E_l$ is a Landau level, the corresponding eigenvalue of the DTQW is $\mu_l = \exp( - i \epsilon E_l)$.

\subsection{Energy-eigenstates}

\subsubsection{Zeroth-order energy-eigenstates: relativistic Landau levels}

\begin{figure}[h!]
    \begin{minipage}[b]{0.47\linewidth}
        \centering \includegraphics[width=7.7cm]{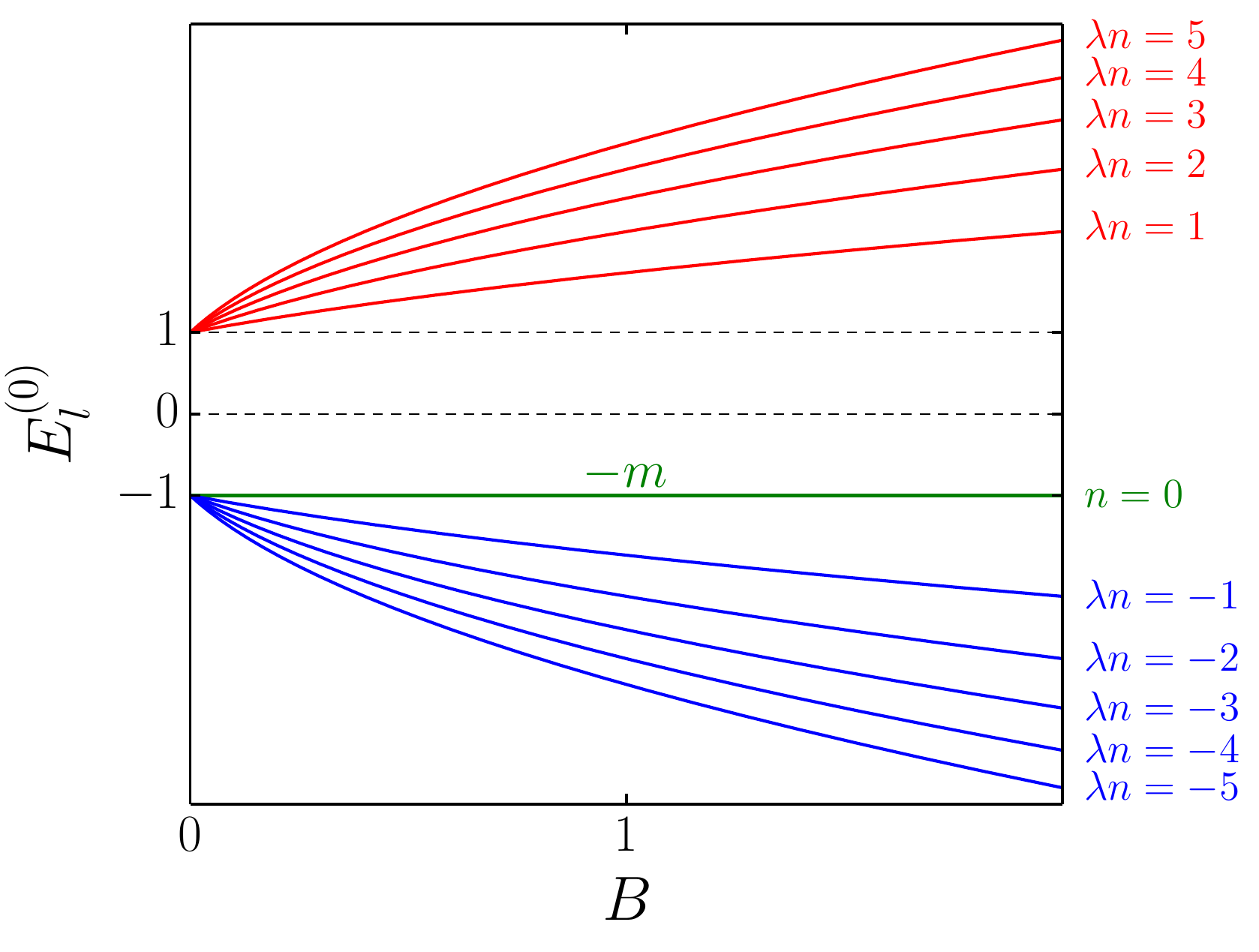}
        \vspace{-0.8cm}
        \caption{Relativistic Landau levels (RLL) $E_l^{(0)}$ as functions of the magnetic field $B$, for a Dirac fermion of mass $m=1$ and charge $q=-1$. The ground state is labelled by $l=n=0$ and the excited states by $l=(\lambda,n)$, with $\lambda = \pm$ and $n \in \mathbb{N}^{\ast}$. \label{Fig:LL_eigenvalues}}
    \end{minipage}\hfill
    \begin{minipage}[b]{0.48\linewidth}
        \centering \includegraphics[width=7.3cm]{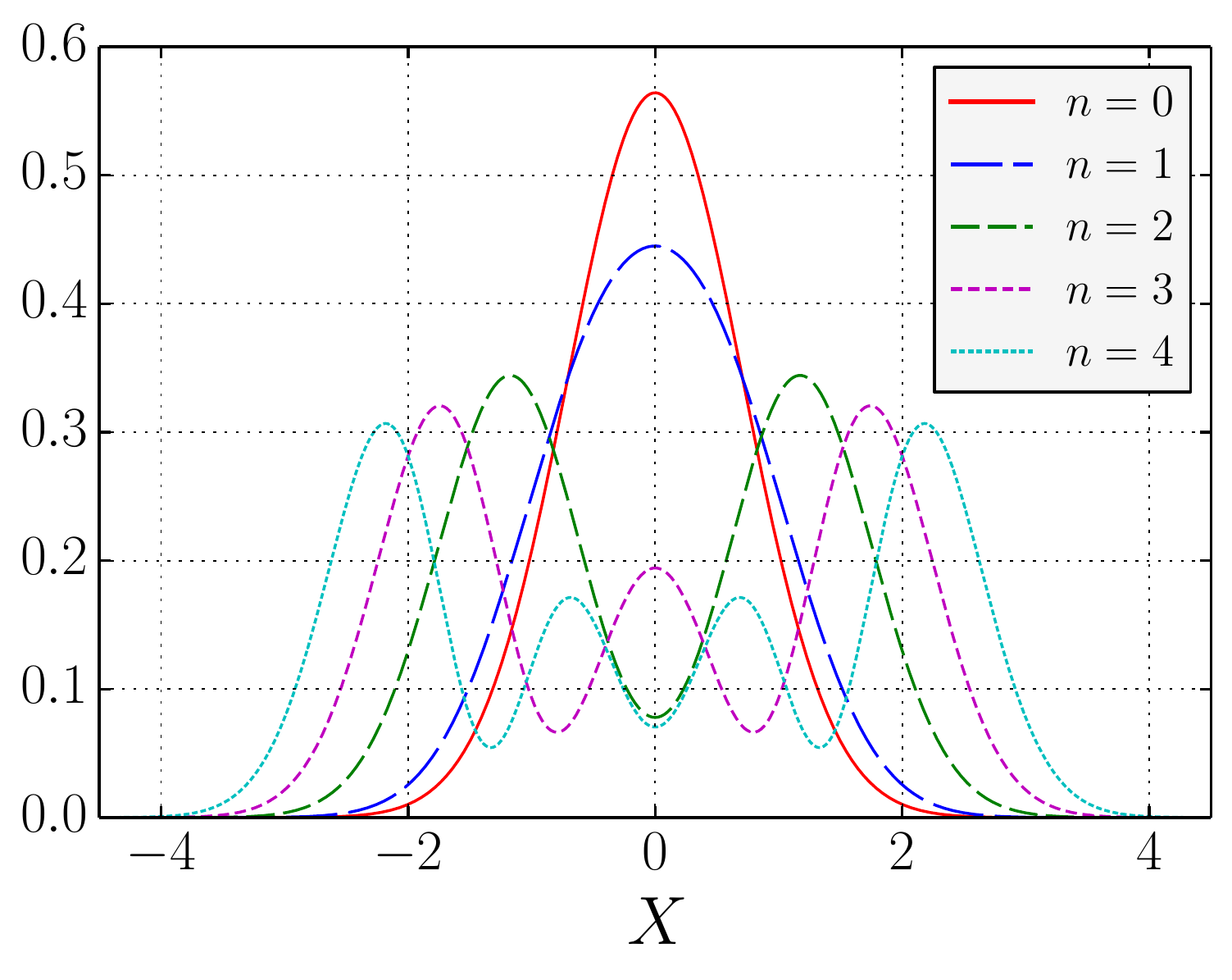}
    \vspace{-0.4cm}
        \caption{Probability densities of the Hermite eigenvectors associated to the relativistic Landau levels, as functions of $X$, for mode $K=0$, magnetic field $B=1$, mass $m=1$ and charge $q=-1$. We have plotted the states $l=n=0$ and $l=(+,n)$ for $n=1,4$. \label{Fig:LL_eigenstates}}
      \vspace{-0.09cm}  
    \end{minipage}
\end{figure}

The zeroth-order Hamiltonian $\mathcal{H}^{(0)} (B, m)$ is 
the Hamiltonian of the Dirac equation (\ref{eq:Dirac}). Its eigenvalues 
are the so-called relativistic Landau levels $E_l^{(0)}$ \cite{Goerbig_review, Bernd_Thaller}. The ground state is labelled by $l=n=0$, and the excited states by $l=(\lambda,n)$, where $\lambda=\pm$ and $n \in \mathbb{N}^{\ast}$. The associated energies are:
\begin{equation}
E_l^{(0)} =  \left\{
\begin{array}{l l l}
- \mbox{sgn} (B) {m} &  \ \ \ l=n=0 \ & \ \ \ \text{ground state} \\
 \lambda \sqrt{m^2 + 2|B|n} & \ \ \ l = (\lambda,n) & \ \ \ \text{excited states} \ .
\end{array}
\right.
\end{equation}
The ground-state energy is proportional to the mass of the fermion, as expected for relativistic `particles'. This energy has the same sign as $(-1 \times B)$, or $qB$ for a generic charge $q$. The sign $\lambda$ of the excited-states energies can be positive or negative, which is expected for Dirac fermions. In the context of high-energy particle physics, the states of negative energy are interpreted as fermions of opposite charge with positive energy. In the context of condensed-matter systems, the states of negative energy correspond to `holes' in the valence band, and the mass corresponds to a `gap' between the valence and conduction bands\footnote{The concepts of `holes' and `mass gap' are general to condensed-matter systems, and not specific to Dirac materials.}, which translates into insulating properties of the material (in the case of graphene, this gap vanishes). The energies of the relativistic Landau levels have a square-root dependence in $|B|n$ (see Fig. \ref{Fig:LL_eigenvalues}), in constrast with the linear dependence of the non-relativistic Landau levels.

Let us now discuss the eigenstates of $\mathcal{H}^{(0)} (B, m)$. To each energy level corresponds an eigenspace of inifinite dimension. There are two usual choices of basis in these eigenspaces. The first basis is made of vectors which are not only eigenvectors of $\mathcal{H}^{(0)} (B, m)$, but also of the $Y$ component $\mathcal{P}_Y$ of the momentum operator. Note that this operator generates translations in the $Y$ direction and that the mesh on which the DTQW is defined is invariant by such translations. This basis is the one used in the present article. It is best determined by working in the representation introduced in Section \ref{ssec:Ham} above and by searching for the eigenvectors of the matrix 
appearing in (\ref{eq:Hamiltonian_form_TF_final}).


The full computation of the vectors constituting this basis is given in Appendix A.
The two spin-components $(\phi^{\pm})_l^{(0)}(X,K)$ of these vectors are, as a function of $X$, two consecutive eigenfunctions of an effective $1D$ harmonic oscillator (HO) centered at point $\chi(B,K) = -K/B$. The fact that $|\chi(B,K)|$ decreases with $|B|$ reflects the confining properties of $B$. The  HO has pulsation $\omega = 2|B|$, which is linear in $|B|$ as the classical cyclotron pulsation. The HO-eigenfunctions have a characteristic length $a = 1/\sqrt{|B|}$, which again traces the confining properties of $B$. Regarding the $n$-dependence of the HO-eigenfunctions, their spatial extension scales as $\sqrt{n}$. Since the $n$-th eigenfunction has $n$ nodes which can roughly be considered equally spaced \cite{CCTannoudji}, these nodes are typically separated by a distance $\sqrt{n}/n = 1 / \sqrt{n}$ . Figure \ref{Fig:LL_eigenstates}
displays, for $K = 0$, 
the probability densities $P_l^{(0)}(X,K)=|(\phi^{-})_l^{(0)}(X,K)|^2 + |(\phi^{+})_l^{(0)}(X,K)|^2$ associated to the first 
eigenvectors.

The second usual choice is to use basis vectors which are eigenstates of both $\mathcal{H}^{(0)} (B, m)$ and of a certain gauge-invariant angular momentum \cite{HRR93}. This choice of basis is not natural in the context of this article, because the angular momentum generates rotations, which are not symmetries of the mesh on which the DTQW is defined.

\subsubsection{First-order energy-eigenstates}

As discussed in the Introduction, Landau levels play a crucial role in many condensed matter phenomena, for example the quantum Hall effect. If one hopes to simulate these phenomena with DTQWs, it is thus important to check that Landau levels exist, not only for the continuous limit, but also for the discrete walks themselves. The purpose of this Section is to provide a perturbative contruction of the Landau levels for the DTQWs at first order in the finite time- and position-step $\epsilon$.



\begin{figure}[!h]
\centering
\includegraphics[width=9cm]{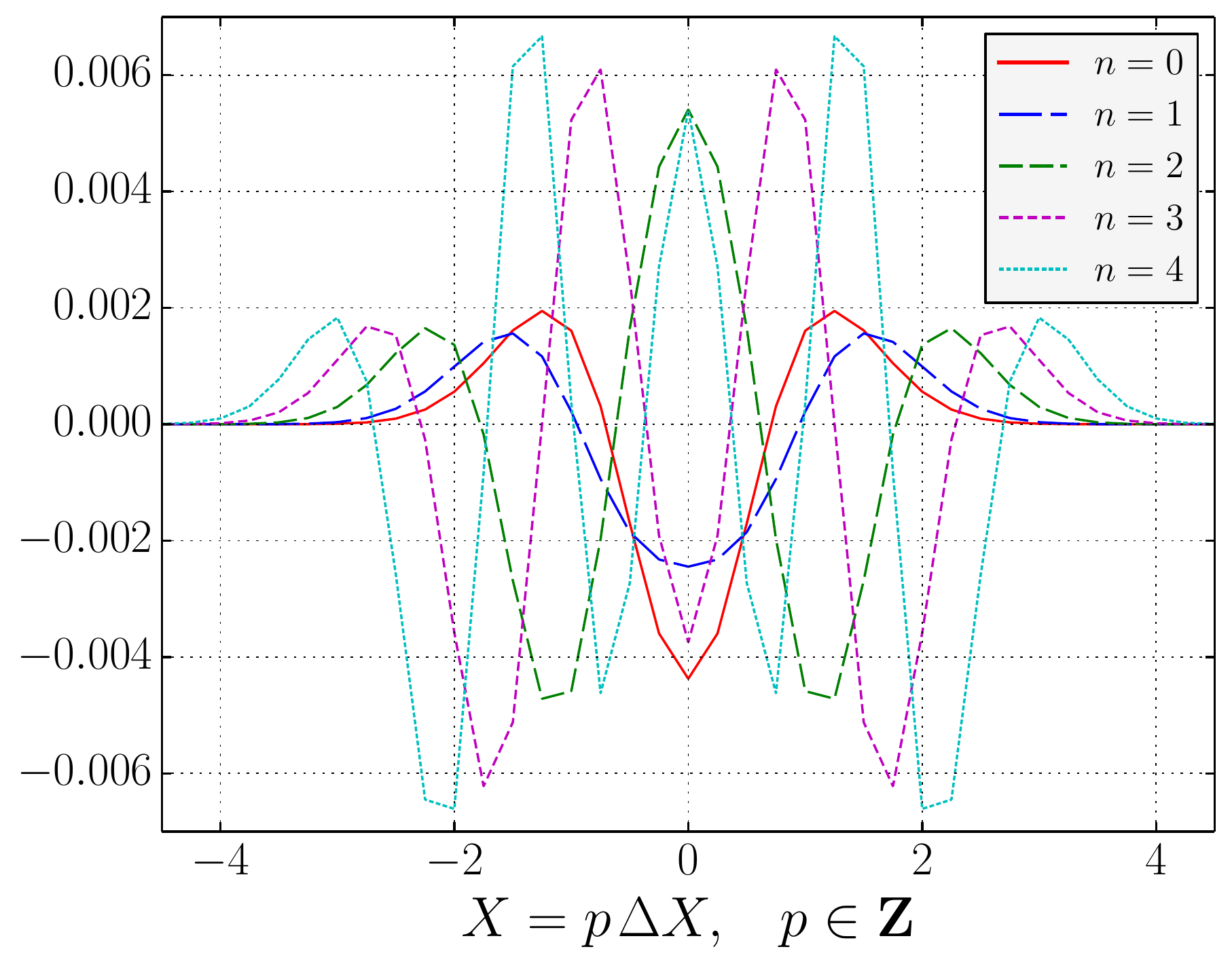}
\vspace{-0.3cm}
\caption{\small 
Differences between the probability densities associated to the Landau levels of the DTQW, computed analytically at first order in the time- and position-step (see Appendix \ref{appendix_order1}), and those associated to the standard relativistic Landau levels. All curves correspond to $\Delta X = 0.25$, $K = 0$, $\lambda_n = +$, $B = 1$, $m = 1$, and $n$ labels the energy state.}
\label{delta_P}
\end{figure}

The eigenvalues and eigenstates of the full Hamiltonian $ \mathbf{H}(\epsilon, B, m)$ can be obtained from those of $\mathcal{H}^{(0)} (B, m)$ by using perturbation theory. 
Appendix B presents the computation at first order in $\epsilon$. It is found that the first-order corrections to the energy levels identically vanish. However, the first-order corrections to the eigenfunctions do not vanish. The probability density of the energy eigenstates of the DTQW is thus different from the probability density of the relativistic Landau levels corresponding to the continuous Dirac dynamics. Typical results are presented in Figure \ref{delta_P}.

All displayed profiles correspond to the Fourier mode $K= 0$ and are thus symmetric with respect to $X = 0$. Giving $\epsilon$ a finite value increases ({\sl resp.} decreases) the central density for levels corresponding to even ({\sl resp.} uneven) values of $n$, the only exception being the fundamental $n = 0$, for which the central density decreases. An $n$-dependent number of maxima and minima also occur on either side of $X = 0$.

The validity of the analytical perturbative computation of the Landau levels for the DTQW can be supported by numerical simulations. 
 {\sl A priori}, the Hamiltonian $\mathbf{H}^{(r)}(\epsilon, B, m) = \sum_{k=0}^{r} \epsilon^k \mathcal{H}^{(k)}(B, m) $ describes the dynamics of the DTQW, not at order $r$, but at order $r+1$ in $\epsilon$. This is so because the Dirac equation (\ref{eq:Dirac}) and the corresponding Hamiltonian (\ref{eq:Hamiltonian_form_TF_final}) are obtained from the exact equations defining the DTQW by a Taylor expansion at first order in $\epsilon$, and thus describe the dynamics of the DTQW, not at zeroth order, but at first order in $\epsilon$ (at zeroth order in $\epsilon$, the DTQW leaves the state unchanged, see above Section \ref{section:continuous_limit}). Let now $E_l^{(r)}$ and 
$\Phi_{l}^{(r)}(X, K)$ be the $l$-th eigenvalue and the corresponding eigenstate of $ \mathbf{H}^{(r)}(\epsilon, B, m)$, both computed at order $r$ in $\epsilon$.
If the perturbative computation is correct, the time-evolution of $\Phi_{l}^{(r)}(X, K)$ by the DTQW should be entirely controlled, at order $r+1$ in $\epsilon$, by $E_l^{(r)}$. In particular, after one time-step of the exact DTQW, $\Phi_{l}^{(r)}(X, K)$ should evolve into a state $W_l^{(r)} (X, K)$ which can be approximated, at order $r + 1$ in $\epsilon$, by ${\tilde W}_l^{(r)} (X, K)= e^{-i E_l^{(r)}  \times \Delta T} \Phi_{l}^{(r)}(X,K)$. For each $K$, the distance between the two functions of $X$, ${W}_l^{(r)} (X, K)$ and 
${\tilde W}_l^{(r)} (X, K)$, can be evaluated by
\begin{equation} \label{eq:relative difference}
\delta_l^{(r)}(K) \equiv \frac{\| W_l^{(r)}(\cdot, K) - {\tilde W}_l^{(r)}(\cdot, K) \|}{\| {\tilde W}_l^{(r)}(\cdot, K) \|} \ ,
\end{equation}
where $\| \cdot \|$ stands for the $L^2$ norm of a position ($X$-)dependent function $\Psi$ defined on the lattice:
\begin{equation} \label{eq:L2_norm}
\| \Psi \| = \left[ \sum\limits_{\substack{p = - p_{\mathrm{max}}(\epsilon)}}^{p_{\mathrm{max}}(\epsilon)}  \left( |\psi^{L}(X_p=p \epsilon)|^2 + |\psi^{R}(X_p=p\epsilon)|^2 \right)  \epsilon \right]^{\frac{1}{2}} ,
\end{equation}
where $p_{\mathrm{max}}(\epsilon)$ scales as $1/\epsilon$.

\begin{figure}[!h]
\begin{center}
\includegraphics[width=9cm]{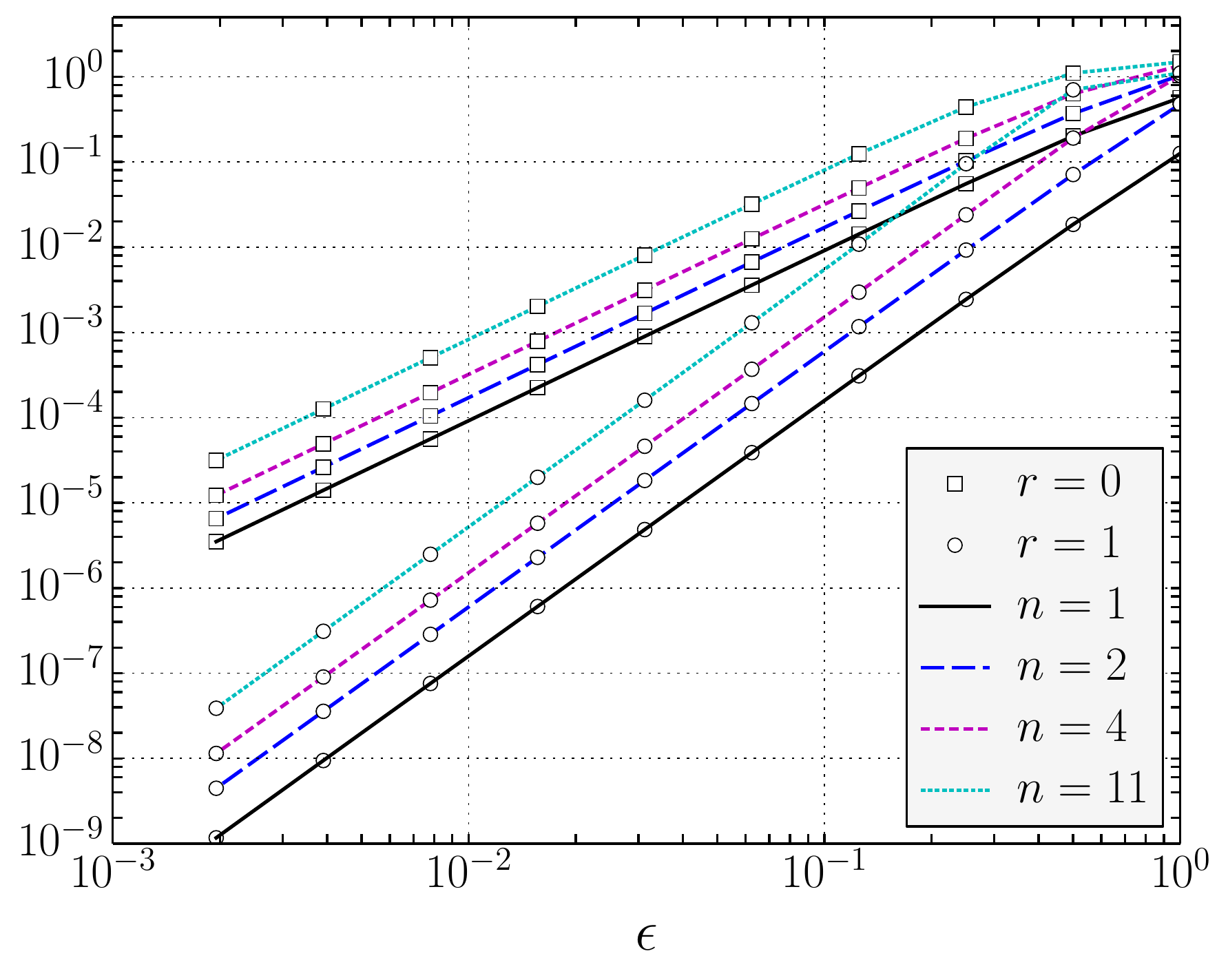}
\vspace{-0.3cm}
\caption{\small Distances $\delta_l^{(r = 0)}(K = 0)$ and $\delta_l^{(r = 1)}(K = 0)$ as functions of $\epsilon$ for various values of $l = (+,n)$, with $B=1$ and $m=1$.}
\label{varying_level}
\end{center}
\end{figure}

\begin{figure}[!h]
\begin{center}
\includegraphics[width=9cm]{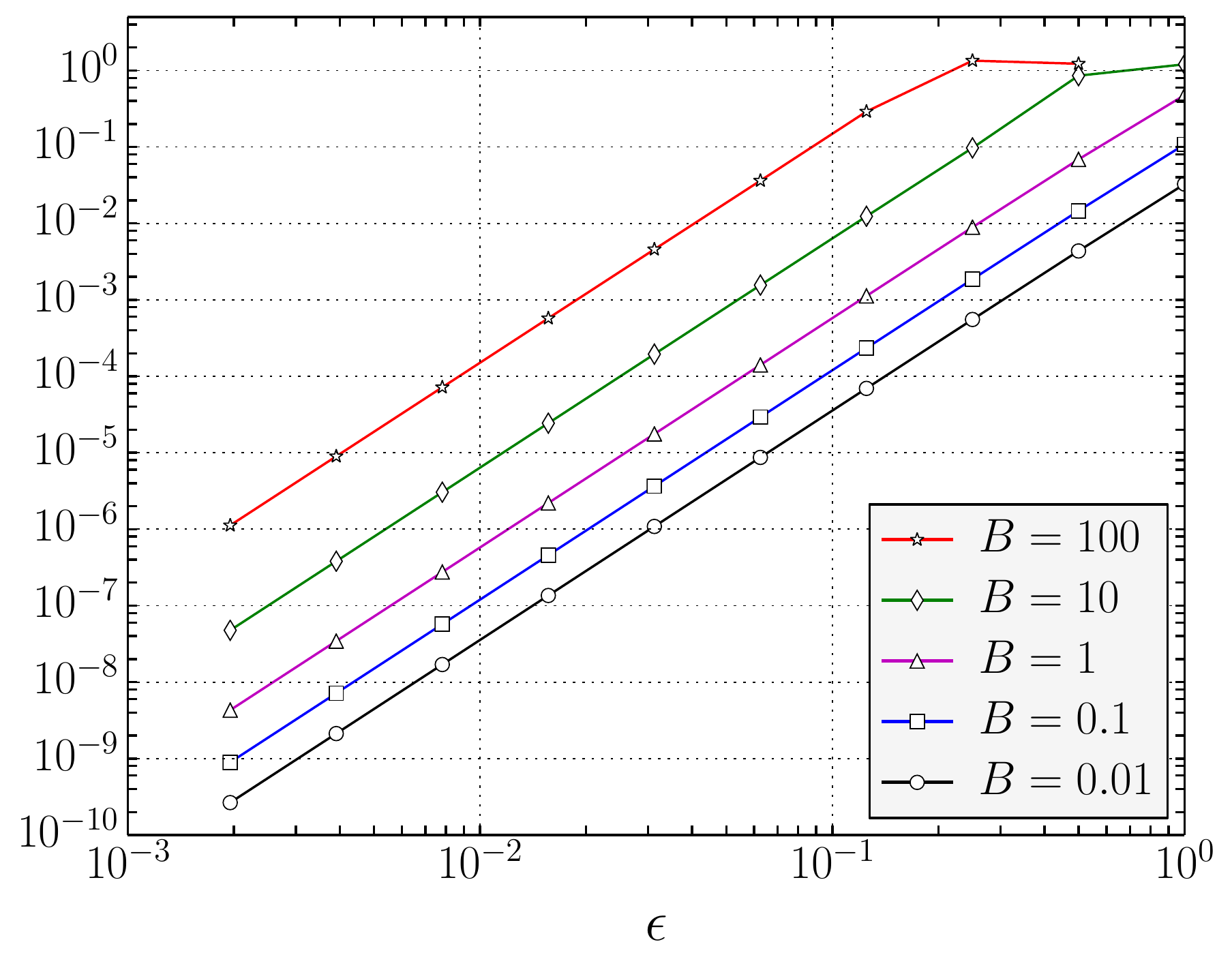}
\vspace{-0.3cm}
\caption{\small Distance $\delta_l^{(r = 1)}(K = 0)$ as a function of $\epsilon$ for various values of $B$, with $l= 0$ and $m = 1$.}
\label{varying_B}
\end{center}
\end{figure}

Figures \ref{varying_level} and \ref{varying_B} display how $\delta_l^{(r = 0)}(K = 0)$ and $\delta_l^{(r = 1)}(K = 0)$ scale with $\epsilon$ for various values of $l = (+,n)$ and for various values of $B$. These figures clearly confirm that $\delta_l^{(r = 0)}(K = 0)$ scales as $\epsilon^2$ and $\delta_l^{(r = 1)}(K = 0)$ scales as $\epsilon^3$ for a large range of $\epsilon$-values which extends well above $10^{-1}$. This numerical result thus supports the litteral perturbative construction of the Landau levels for the DTQW.

\vspace{0.2cm} 
 
\newpage

\section{Non stationnary density profiles of the DTQW}


Let us now look at some interesting non-stationnary density profiles generated by the DTQW dynamics. We will consider simple localized initial states and the corresponding trajectories. These are clearly outside the scope of the continuous limit and provide an outlook of the phenomenolgy one might expect from an experimental realisation of  the $2D$ DTQWs under consideration. These simulations support the interpretation of the parameter $B$ as a magnetic field even outside the continuous limit.
In what follows, the time-and space-step $\epsilon$ is set to unity so that the instants have coordinates $T_j = j \epsilon = j$ and the vertices of the mesh have coordinates $X_p = p \epsilon = p$ and $Y_q = q\epsilon = q$ with $(j, p, q) \in {\mathbb N} \times {\mathbb Z}^2$. The probability `density' of the walker at time $j$ and point $(p, q)$ is defined by $P (j, p, q) = |\psi^L(j, p, q)|^2 + |\psi^R(j, p, q)|^2 \ .$

We consider
%
initial states of Gaussian probability density 
\begin{align}
G_w(p, q) =  
\frac{1}{\mathcal{N}_w} N_w(p, q) =
\frac{1}{\mathcal{N}_w} \frac{e^{-\frac{p^2+q^2}{2w^2}}}{(\sqrt{2 \pi} w)^2} \ ,
\end{align}
where $w$ is a strictly positive real number and 
$\mathcal{N}_w$ is the normalization constant
on the $2D$-lattice:
\begin{equation}
\mathcal{N}_w = \sum_{p=-p_{\mathrm{max}}}^{p_{\mathrm{max}}} \ 
    \sum_{q=-q_{\mathrm{max}}}^{q_{\mathrm{max}}}
     N_w(p, q)  \ .
\end{equation}
By convention, $G_0$ is the Dirac mass at the origin of coordinates:
\begin{equation}
G_{w=0}(p, q) = \left\{
\begin{array}{c l}
1 & \mathrm{if} \ (p ,q)=0 \\
0 & \mathrm{elsewhere}
\end{array}
\right. \ .
\end{equation} 
Of course, choosing an initial density does not fully specify the initial wave functions $\psi^{L/R}(j=0, p ,q)$. 
We choose the following initial condition, for which the left component of the walker is a real number and carries all the probability density:
\begin{align}\label{eq:initcondw}
\psi^{L}_w(j=0,p ,q) & = \sqrt{G_w(p, q)} \\ \nonumber
\psi^{R}_w(j=0,p ,q) & = 0 \ .
\end{align}


The $p$-spread of the distribution at time $j$ is defined by:
\begin{equation}
\sigma^p_w(j) = \left[ \sum_{p=-p_{\mathrm{max}}}^{p_{\mathrm{max}}} \ 
    \sum_{q=-q_{\mathrm{max}}}^{q_{\mathrm{max}}}
     p^2 \, P_w(j, p, q) \right]^{\frac{1}{2}} \ . 
\end{equation}
where $P_w$ is the probability density generated by the initial condition (\ref{eq:initcondw}). The $q$-spread is defined in a similar manner.

\begin{figure}[!h]
\begin{center}
\includegraphics[width=9cm]{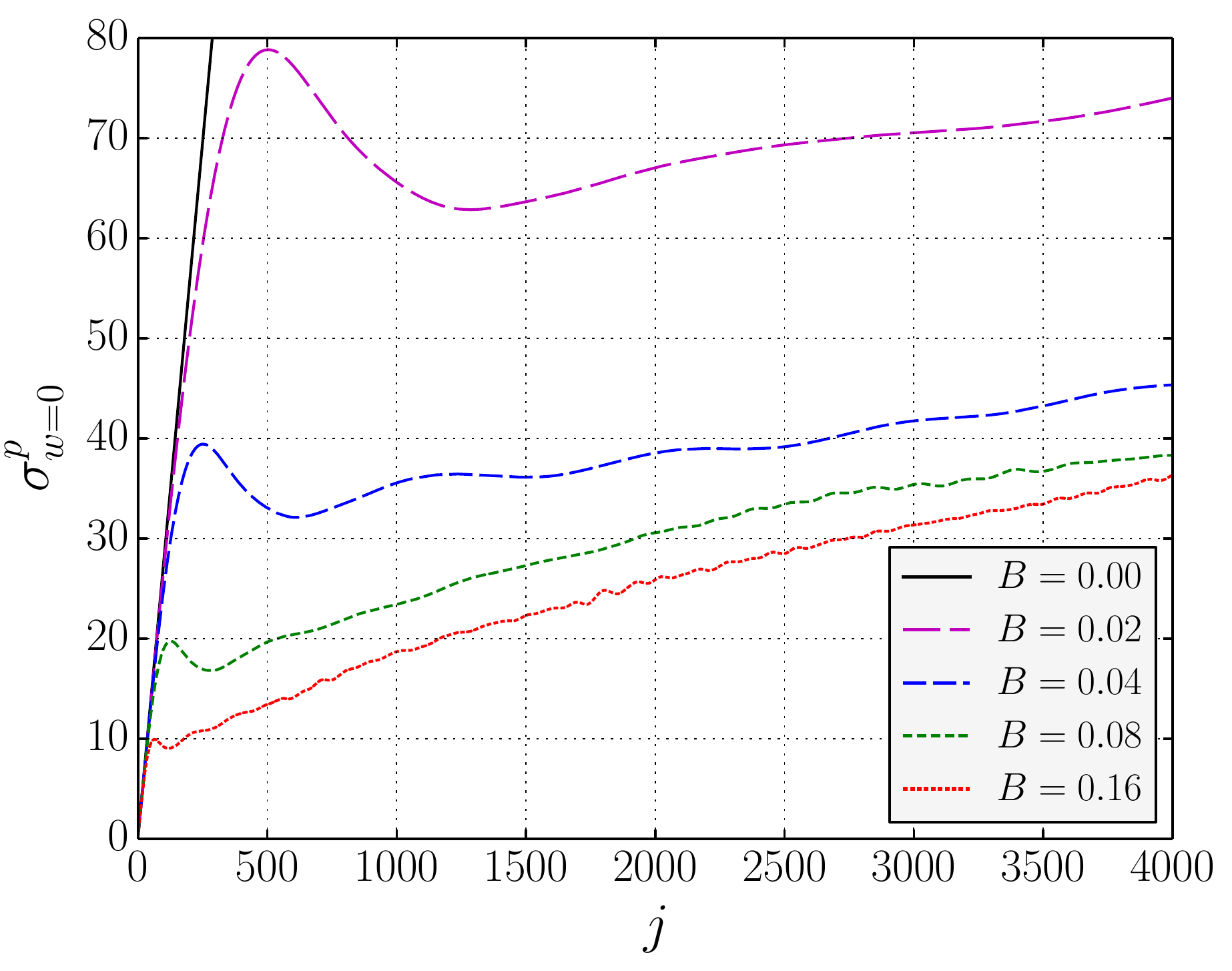}
\vspace{-0.3cm}
\caption{\small Time evolution of the $p$-spread of a Dirac-delta initial condition evolved through our $2D$ DTQW with $m=1$, for different values of $B$.}
\label{fig:X_spread}
\end{center}
\end{figure}

Figure \ref{fig:X_spread} displays the time evolution of the $p$-spread 
for $m=1$ and different values of $B$. For all values of $B$ considered in this figure, the $q$-spread is equal to the $p$-spread up to the precision of the numerical computation (data not shown).
This figure shows that the walk with $B=0$ propagates ballistically. For non vanishing $B$, the dynamics is at first ballistic and then followed by a quasi-confined regime whose radius is, at a given time, a decreasing function of $B$. Note also that the time during which the DTQW behaves ballistically decreases as $B$ increases.
  
\begin{figure}[!h]
\begin{center}
\includegraphics[width=5.95cm]{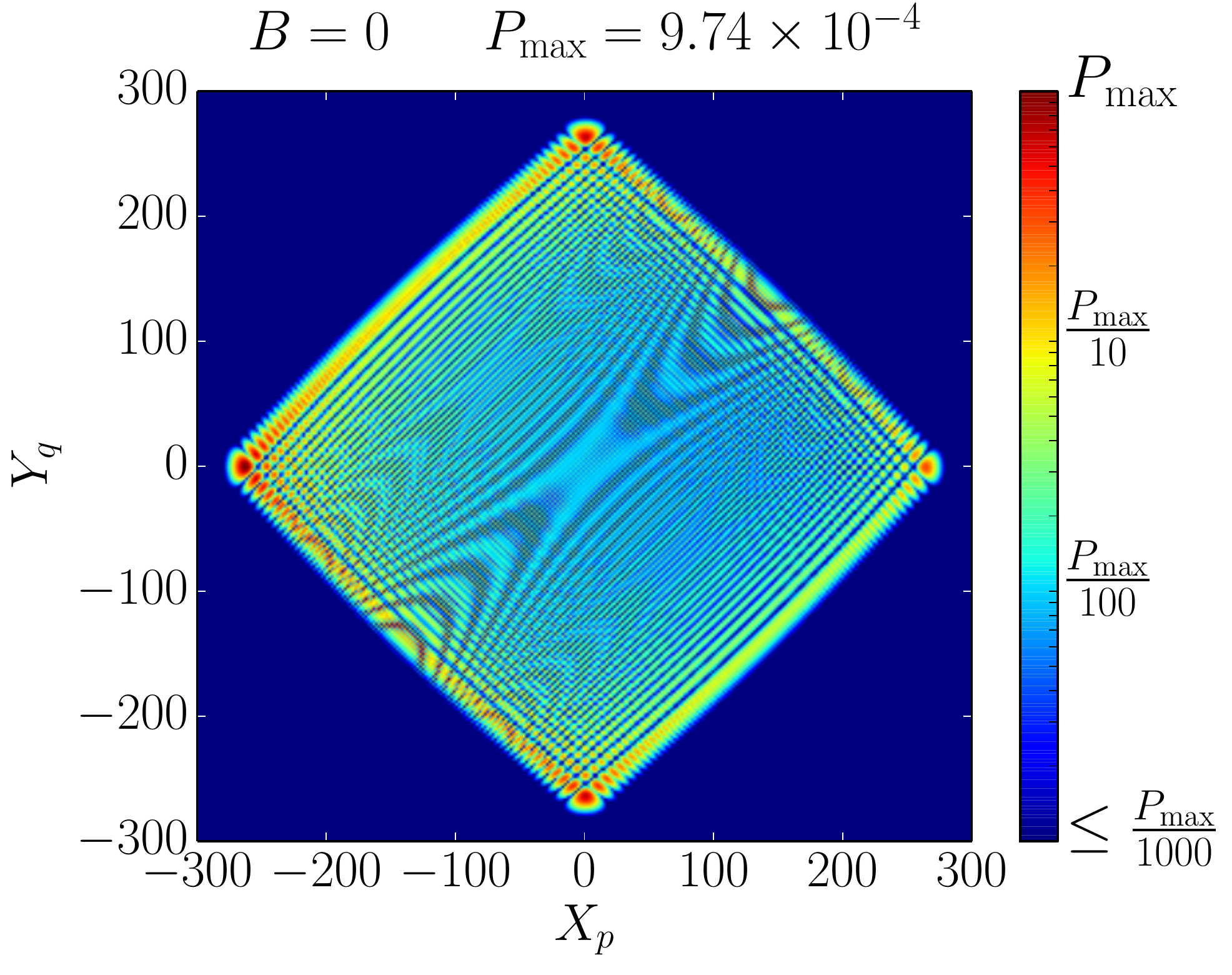}
\includegraphics[width=5.18cm]{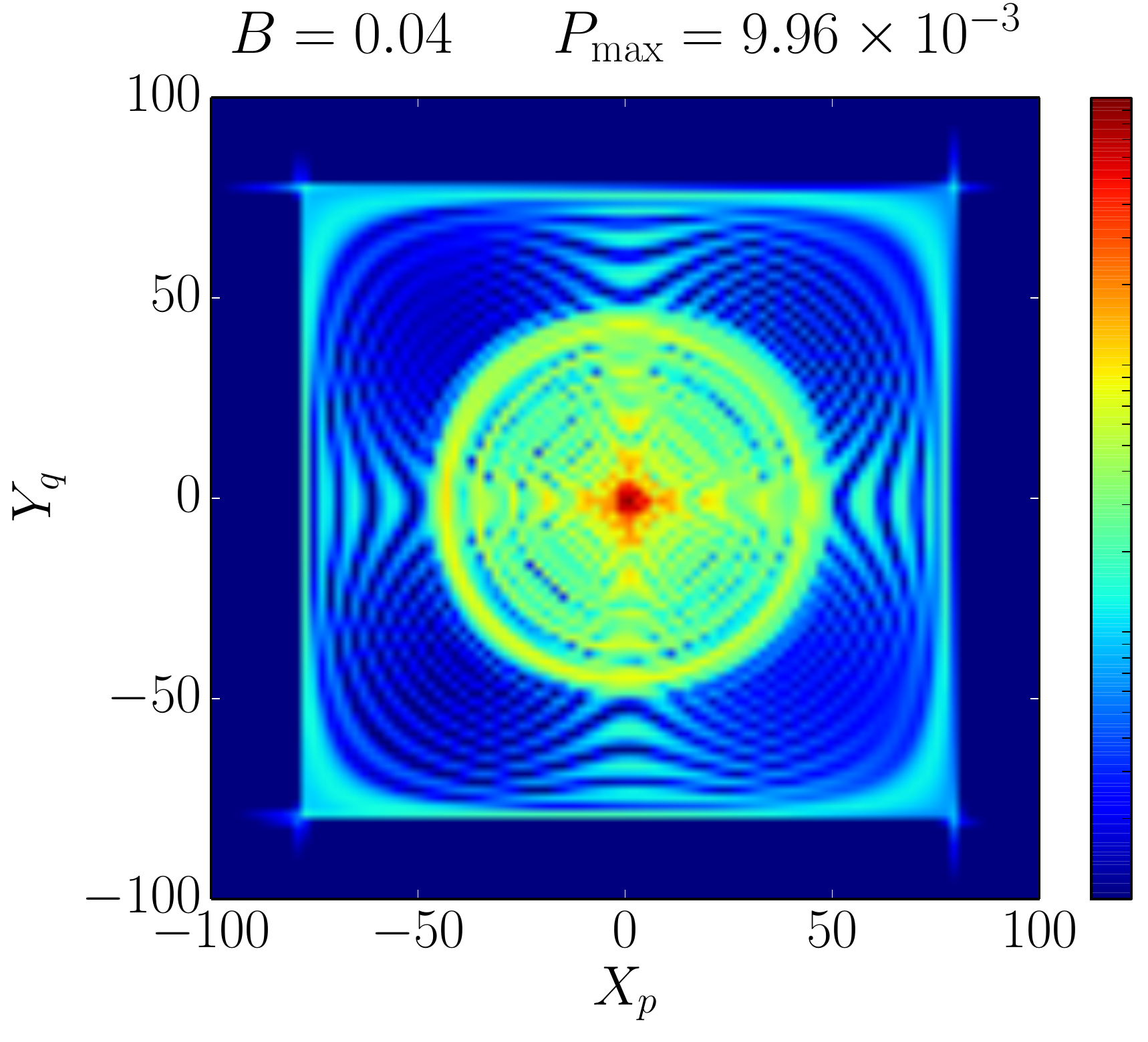}
\includegraphics[width=5.05cm]{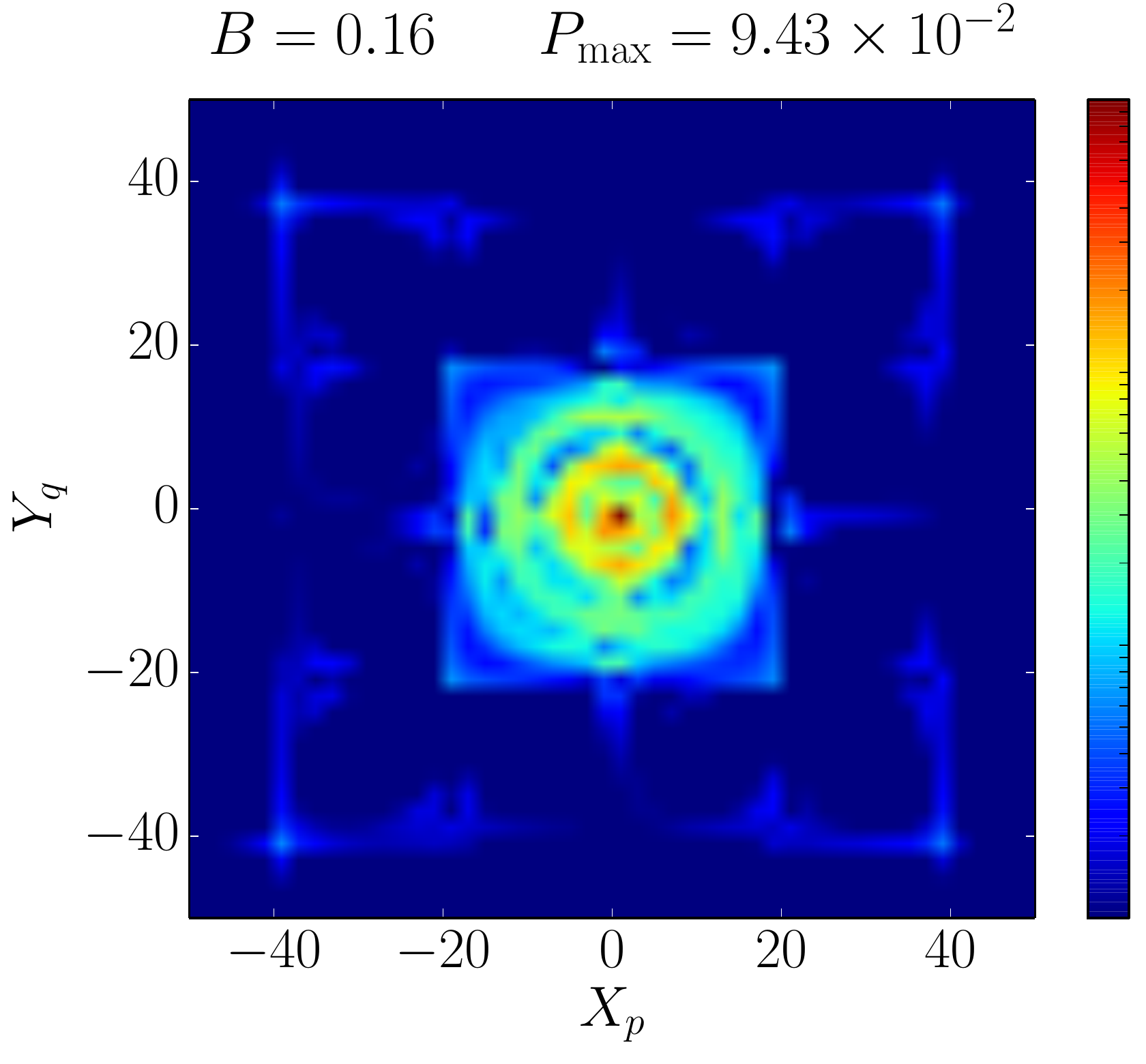}
\vspace{-0.3cm}
\caption{\small Probability density of Dirac-delta initial condition evolved through our $2D$ DTQW with $m=1$, for different values of $B$, at time $j=500$. We show this probability density over 3 decades, from its maximum value $P_{\max}$ down to $P_{\max}/1000$. The initial state having a Dirac-delta probability density carried by a single spin component, the probability density vanishes every 2 steps in both $X$ and $Y$ directions; to have a better visualization of the patterns, we have only plotted the non-vanishing values and (cubically) interpolated them.}
\label{fig:typical_dynamics_varying_B}
\end{center}
\end{figure}

     Figure \ref{fig:typical_dynamics_varying_B} shows, for $m=1$ and different values of $B$, the  probability density contours at time $j = 500$.
The zero-$B$ plot (left) shows a maximum located on the left front, which is about $3$ times bigger than the right-front local maximum ($3.08 \times 10^{-4}$), while the top- and down-front local maxima are almost equal ($6 \times 10^{-4}$). Increasing $B$ localises the walker in a roughly circularly-symmetric profile within a radius which is (again) a decreasing function of $B$. Taken together, Figures \ref{fig:X_spread} and \ref{fig:typical_dynamics_varying_B} show that the parameter $B$ has on the DTQW the same qualitative effects as those induced by a standard magnetic field on a Dirac fermion moving in continuous space-time. Note that both figures display results which fall well outside the scope of 
the continuous limit.

\vspace{0.0cm}

\begin{figure}[!h]
\begin{center}
\includegraphics[width=5.95cm]{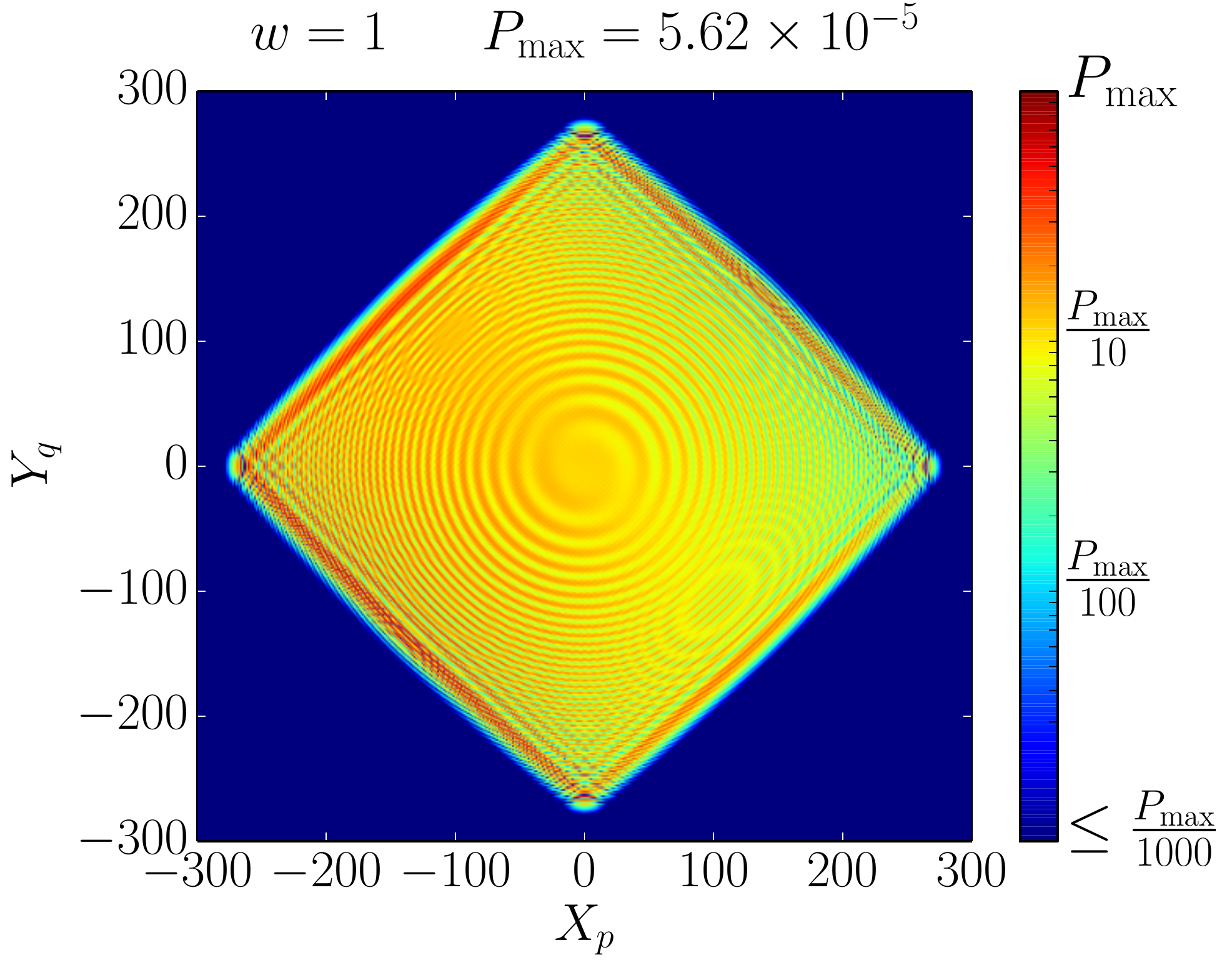}
\includegraphics[width=5.15cm]{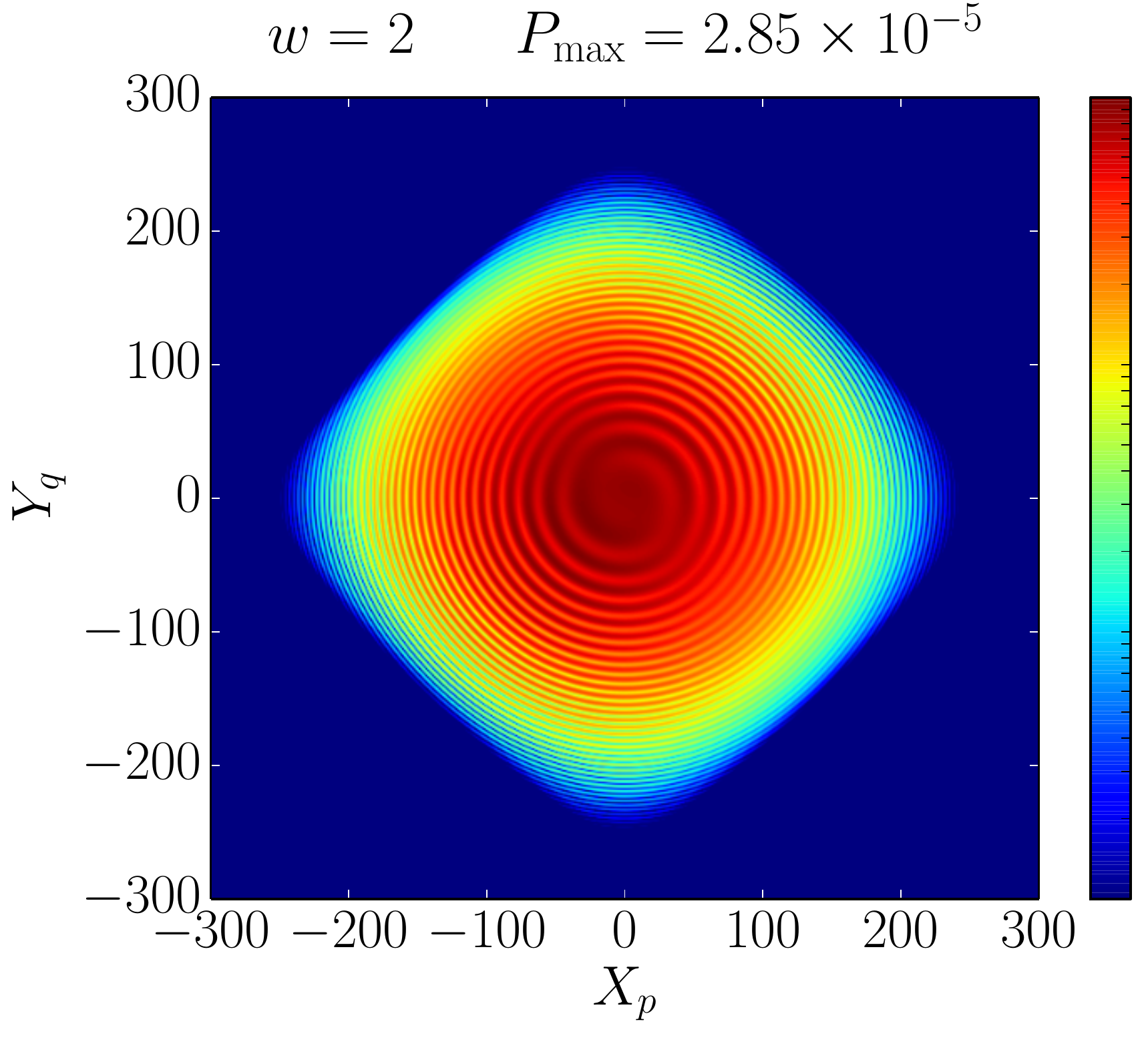}
\includegraphics[width=5.15cm]{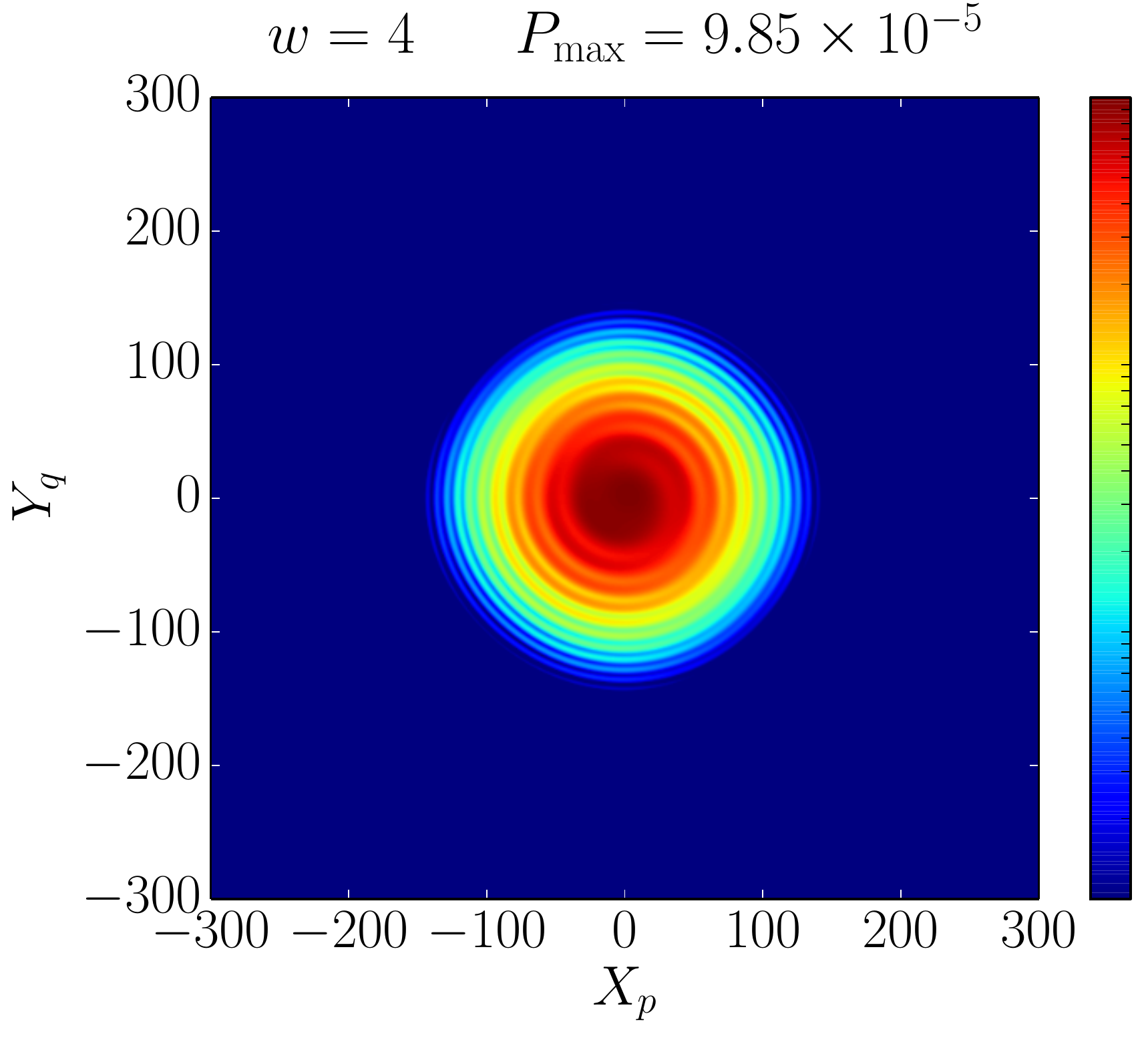}
\vspace{-0.3cm}
\caption{\small Probability density, at time $j=500$, of initial states of increasing width $w$ from left to right, evolved through our $2D$ DTQW with $m=1$ and $B=0$.}
\label{fig:typical_dynamics_varying_w}
\end{center}
\end{figure}

    Figure \ref{fig:typical_dynamics_varying_w} displays contours of the probability density $P_w$ for different values of $w$. The time is $j = 500$, the mass is equal to unity and the magnetic field vanishes. As $w$ increases, the interference patterns gets more and more circular, and the presence of the underlying lattice becomes less and less noticeable.



\newpage

\section{Discussion}

We have constructed a family of $2D$ DTQWs which admits as continuous limit the standard Dirac dynamics of a relativistic spin 1/2 fermion coupled to a constant and uniform magnetic field. We have then shown pertubatively that Landau levels exist for this family, not only in the continuous limit, but also for finite non-vanishing values of the time- and position-step.
We have finally presented numerical simulations which (i) support the existence of Landau levels for the DTQWs (ii) suggest that the magnetic interpretation of the discrete dynamics is still valid even outside the continuous limit.
%
%
%

Let us now discuss the results presented in this article. The family of $2D$ DTQWs considered in this work lives in a two-dimensional Hilbert space {\sl i.e.} the wave function of the walk has two components. This is consistent with standard spinor theory, because irreducible representations of the Clifford algebra in $(2+1)$ dimensional space-times are indeed two-component spinors. 
To explore the full $2D$ lattice, each walk in the family alternates between time-steps in which it propagates in the $X$-direction and time-steps in which it propagates in the $Y$-direction.
A general walk exploring the $2D$ lattice in this way is defined by two $2 \times 2$ matrices belonging to $SU(2)$ and a phase. Each of the two matrices is caracterized by three Euler angles. Thus, a general $2D$ DTQW with two components is defined by seven real parameters (angles), which generally depend on time and space. For the record, a general $1D$ DTQW with two components is defined by an arbitray time- and space-dependent matrix in $SU(2)$ and a phase {\sl i.e.} by four time- and space-dependent parameters (angles). Previous investigations of $1D$ DTQWs coupled to artificial electric fields teach that the mass of the walk is encoded in the `$\theta$' Euler angle and that electric fields are encoded in the phase parameter and in one of the remaining two Euler angles. This experience has been used in constructing the DTQWs studied in this article.

Building the family (\ref{eq:fund}) of DTQWs also required a choice of gauge. We have chosen to work in a gauge where the vector potential depends only on one cartesian coordinate ($X$), because this makes it possible to use Fourier transforms with respect to the other cartesian coordinate ($Y$), which facilitates the computation of Landau levels. All numerical computations have been carried out in $(X, K)$-space, where $K$ is the wave number associated to $Y$. In particular, no Fourier transform with respect to $X$ has been used.

Generally speaking, all results pertaining to the continuous limits of DTQWs can be envisaged from two different points of view. The first one consists in viewing the DTQWs as discrete schemes approximating partial differential equations. The second one consists in viewing the DTQWs as physical systems in their own right, and to consider their continuous limits as tools allowing a greater insight into a fundamentally discrete dynamics. This second point of view appears especially natural in the context of quantum algorithmics, and it has been strengthened by the existence of recent laboratory experiments on DTQWs. This is the point of view we have adopted in the present article. Let us nevertheless discuss rapidly our results from the first point of view. 

The Dirac equation is notoriously difficult to discretize \cite{FGLB12a,VLK14a}. Naive symmetric discretizations suffer from the so-called fermion-doubling phenomenon \cite{Tref82a}. Other methods are based on the Wilson \cite{Kogut83a} or staggered \cite{KogSuss75a,Susskind77a,Kogut83a} fermion discretizations, but they break chiral symmetry and they also exhibit instabilities and spurious diffusive behaviour. A particularly efficient method has been proposed in \cite{FGLB12a,FGLB14a}. The DTQWs defined by (\ref{eq:fund2}) are symmetric discretization. As such, they do exhibit fermion doubling, and this can be checked for example by a direct examination of the dispersion relation  associated to (\ref{eq:fund2}). Thus, DTQWs are not the best tool to compute accurately numerical solutions of the Dirac equation. This however does not invalidate the use of DTQWs as quantum simulators of condensed-matter physics. Indeed, as already discussed in Section \ref{section:continuous_limit}, the Dirac equation is used in condensed matter only as an effective description of small wave number excitations and the description of excitations in terms of Dirac fermions moving in continuous space certainly becomes invalid at wave numbers comparable to the size of the Brillouin zone of the material, where the excitations `see' the size of the underlying lattice. For small wave numbers, the DTQWs do reproduce the Dirac equation. Because of fermion doubling, the DTQWs do {\sl not} reproduce the Dirac equation for wave numbers comparable to the size of the Brillouin zone but for these wave numbers, the Dirac equation is not physically realistic anyway. Thus, fermion doubling does not preclude the use of DTQWs as quantum simulators of small wave number excitation transport in condensed matter. Moreover, since DTQWs are based on a lattice, they are probably the best tool to model, in graphene-like materials, the transport of excitations with wave numbers comparable to the size of the Brillouin zone, which `see' the underlying lattice \cite{Chandra_report_2013,Sarkar_Paul_2015}. Let us finally refer to \cite{SFGP15a} for an in-depth presentation of the links existing between QWs, the Dirac equation and the quantum lattice Boltzmann model.

A final, rather technical remark is in order. Obtaining the Dirac equation (\ref{eq:Dirac}) as the continuous limit of the DTQW considered in this article requires two very different ingredients. The first one is the identification of the indices $j$, $p$ and $q$ with integers labelling points on a regular cartesian lattice in space-time. The second ingredient is the choice of the same step $\epsilon$ for all three space-time directions. Both ingredients are non trivial and changing them could prove useful in certain contexts. For example, choosing the integers $j$, $p$ and $q$ as labels of arbitrary, non necessarily cartesian coordinates could be useful to study DTQWs on manifolds. And having different steps for the different coordinates could help studying solutions of the DTQWs with different variation scales in the time and/or in the two spatial directions. 
 
The conclusions presented in this article are an extension of previous results which show that several families of $1D$ DTQWs can be interpreted as Dirac spin 1/2 fermions coupled to arbitrary non-constant and non-homogeneous electric and gravitational fields. Preliminary computations show that the family considered in this article can be extended into a larger family where the $2D$ quantum walker can be coupled to arbitrary electromagnetic fields as well as general relativistic gravitational fields. This extension will be presented elsewhere. Even larger extensions which include other gauge fields should be investigated too. Note that $1D$ and $2D$ DTQWs based on four-component wave functions have also been proposed \cite{Sarkar_Paul_2015}. The connection of these alternate walks to relativistic fermions coupled to electromagnetic fields certainly deserves exploring.

Finally, the above results strongly suggest that DTQWs can be used for spintronics and for quantum simulation of condensed-matter systems under magnetic fields, including possible simulations of the quantum Hall effect.

\appendix

\section{Relativistic Landau levels} \label{appendix_LL}


\subsection{Normalized eigenfunctions of the squared zeroth-order Hamiltonian}

The first step in searching for the relativistic Landau levels is to search for the eigenvalues and eigenvectors of $ \left( \mathcal{H}^{(0)} (B, m) \right)^2 $ in basis $(b_-,b_+)$, because this operator is diagonal in that basis. From (\ref{eq:Hamiltonian_form_TF_final}), one is thus led to solving the system
\begin{align}
\label{OH_eq_1}
\mathcal{E}^{\pm} \phi^{\pm} & = \left(-\frac{\partial_{XX}}{2 M} + \frac{1}{2}M \omega^2 (X - \chi(K, B))^2 \right) \phi^{\pm} \ ,
\end{align}
\noindent
where $(\phi^{-}, \phi^{+})$ are the two components of some eigenfunction of $\left( \mathcal{H}^{(0)} (B, m)\right)^2_{(X,K)}$ in basis $(b_-,b_+)$, $E^2$ is the corresponding eigenvalue
of $\left( \mathcal{H}^{(0)} (B, m)\right)^2$ and
\begin{align}
M = 1/2 \ , \ \ \ \ \ \ \omega  = 2|B| \ ,  \ \ \ \ \
\mathcal{E}^{\pm} = E^2  - m^2 \pm B.
\end{align}

Suppose first that both $\phi^{-}$ and $\phi^{+}$ do not vanish identically.  Equations (\ref{OH_eq_1}) then imply
\begin{align}
& E^2  =  (n^{\pm} +\frac{1}{2}) \times 2 |B| + m^2 \mp B, 
\label{eq:relativistic_energies} \\ 
& \phi^{\pm}(X, K)  = \phi^{HO}_{n^{\pm}}(X - \chi(K, B)) \ ,
\end{align}
\noindent
where $n^\pm \in \mathbb N$ and 
$\phi^{HO}_{n}$ is the normalized $n$-th $1D$ harmonic oscillator energy eigenstate, which reads
\begin{align}
\phi^{HO}_{n}(X) = \frac{e^{-\frac{X^2}{2 a^2}}}{\pi^{1/4} \sqrt{2^n n!} \sqrt{a}} \  H_n( X/a) \ , \ \ \ \ \ \mathrm{with} \ \ \  \ H_n(X) = (-1)^n e^{X^2} \partial_X^n e^{-X^2}, \ \ \  \ a = \frac{1}{\sqrt{|B|}} \ .
\end{align}
Equations (\ref{eq:relativistic_energies}) lead to $(n^+ - n^-) = \mbox{sgn} (B)$. The eigenvectors of $\left( \mathcal{H}^{(0)} (B, m)\right)^2$ are thus of the form 
\begin{align}
\label{eigenstate_1}
& 
B > 0 \ \mathbf{:} \ \ \ \ 
[(\Pi_{n}^{B>0})^u(X, K)] =
\left(
  \begin{array}{c}
  \alpha \ \phi^{HO}_{n-1}(X - \chi(K, B))\\
  \beta \ \phi^{HO}_{n}(X - \chi(K, B))
  \end{array}
\right) \\
\label{eigenstate_2}
& 
B < 0 \ \mathbf{:} \ \ \ \ 
[(\Pi_{n}^{B<0})^u(X, K)] = 
\left(
  \begin{array}{c}
  \beta \ \phi^{HO}_{n}(X - \chi(K, B))\\
  \alpha \ \phi^{HO}_{n-1}(X - \chi(K, B))
  \end{array}
\right),
\end{align}
where $u \in \{-,+\}$, $n \in \mathbb N^*$ and $\alpha$, $\beta$ are two complex numbers obeying $|\alpha|^2 + |\beta|^2 = 1$ (normalized eigenvectors).

The squared Hamiltonian $\left( \mathcal{H}^{(0)} (B, m)\right)^2$ also admits eigenvectors with one identically vanishing spin component in basis $(b_-,b_+)$. For these eigenvectors, Equations
(\ref{OH_eq_1}) degenerate into a single equation. The non-vanishing spin component, say $\phi^{-}$ ({\sl resp.} $\phi^{+}$) is then identical to an arbitrary eigenfunction $\phi^{HO}_n$ of the harmonic oscillator Hamiltonian and the corresponding eigenvalue of $\left( \mathcal{H}^{(0)} (B, m)\right)^2$ is 
$\varepsilon^2_n = (n + \frac{1}{2}) \times 2|B| + m^2 + B$ ({\sl resp.} $\varepsilon^2_n = (n + \frac{1}{2}) \times 2|B| + m^2 - B$).

\subsection{Normalized eigenfunctions of the zeroth-order Hamiltonian}

Any normalized eigenfunction of the Hamiltonian $\mathcal{H}^{(0)} (B, m)$ is also a normalized eigenfunction of its square and is thus one of the functions determined above. 
The discrimination between the functions that are actual eigenfunctions of $\mathcal{H}^{(0)} (B, m)$ and those which are not is best made by computing directly the action of the Hamiltonian $\mathcal{H}^{(0)} (B, m)$ on all eigenfunctions of $\left( \mathcal{H}^{(0)} (B, m)\right)^2$. 

One thus finds the ground-state energy $E_0^{(0)}= - \mbox{sgn} (B) {m}$ and the corresponding normalized eigenfunction $\Phi_0^{(0)}(X, K) = {\phi}_0^{HO} (X - \chi(K, B)) \ \! b_{\mbox{sgn} (B)}$, which has an identically vanishing spin component. The other eigenstates are labelled by an integer $n \in \mathbb N^{\ast}$ and a sign $\lambda = \pm 1$. The eigenvalues are\\ $E_{\lambda,n}^{(0)} = \lambda \sqrt{m^2 + 2|B|n}$ and the normalized eigenstates, that we note $\Phi_{\lambda,n}^{(0)}(X,K)$, are of the form (\ref{eigenstate_1}), (\ref{eigenstate_2}) with  
\begin{align}\label{eq:alpha_beta}
\alpha_{\lambda,n} = -i \sqrt{\frac{2|B|n}{(E_{\lambda,n} -  {\mbox{sgn} (B)} m)^2 + 2|B|n}} \ , \ \ \ \ \ \ \ \
\beta_{\lambda,n} = \frac{E_{\lambda,n} - {\mbox{sgn} (B)} m}{\sqrt{(E_{\lambda,n} - {\mbox{sgn} (B)} m)^2 + 2|B|n}} \ .
\end{align}

\section{First-order corrections to the eigenvalues and eigenstates of the zeroth-order Hamiltonian} \label{appendix_order1}

As seen in Appendix A, the eigen-elements of $\mathcal{H}^{(0)}(B,m)$ are labelled by $l=0 \ \mbox{or} \ (\lambda,n)$, with $\lambda=\pm$ and $n \in \mathbb N^{\ast}$ \footnote{In this labelling, we forget about the quantum number $K$, which can be considered as a parameter in our problem. Indeed, this quantum number $K$ corresponds to an essential degeneracy of the full Hamiltonian of the DTQW, namely $\mathbf{H}^{(1)}(\epsilon,B,m)$; this degeneracy will hence not be removed by our perturbative computation.}. We now apply perturbation theory for the Hamiltonian $\mathbf{H}^{(1)}(\epsilon,B,m) = \mathcal{H}^{(0)}(B,m) + \epsilon \mathcal{H}^{(1)}(B,m)$. To lighten notations, we define
\begin{equation}
\langle l' | \mathcal{H}^{(1)} | l \rangle \equiv \int_{-\infty}^{+\infty} dX \sum\limits_{\substack{u = \pm \\ v = \pm}}  \left( ( \Phi_{l'}^{(0)}(X,K) )^u \right)^{\ast}  \left. {   (\mathcal{H}^{(1)}(B,m))^{u}}_v \right|_{(X,K)}  ( \Phi_{l}^{(0)}(X,K) )^v \ .
\end{equation}
\noindent
At first order in $\epsilon$, the eigenstates of $\mathbf{H}^{(1)}(\epsilon,B,m)$ can be written $\Phi_l^{(1)} = \Phi_l^{(0)} + \epsilon \Delta^{(1)}_{l}$, with:
\begin{align}
\label{eigenstate1_0}
\Delta^{(1)}_{0}   & = \sum_{\lambda'= \pm 1} 
\sum_{n' \neq 0} \frac{\langle \lambda',n' | \mathcal{H}^{(1)} | 0 \rangle}{E^{(0)}_{0} - E^{(0)}_{\lambda',n'}} \  \Phi^{(0)}_{\lambda',n'}  \ , \\
\label{eigenstate1}
\Delta^{(1)}_{\lambda,n}  & = \sum_{\lambda'= \pm 1} 
\sum\limits_{\substack{n' \neq 0 \\ n' \neq n}} \frac{\langle \lambda',n' | \mathcal{H}^{(1)} | \lambda,n \rangle}{E^{(0)}_{\lambda,n} - E^{(0)}_{\lambda',n'}} \  \Phi^{(0)}_{\lambda',n'}  \ + \ 
\frac{\langle -\lambda,n | \mathcal{H}^{(1)} | \lambda,n \rangle}{E^{(0)}_{\lambda,n} - E^{(0)}_{-\lambda,n}} \  \Phi^{(0)}_{-\lambda,n}  \ + \
\frac{\langle 0 | \mathcal{H}^{(1)} | \lambda,n \rangle}{E^{(0)}_{\lambda,n} - E^{(0)}_{0}} \  \Phi^{(0)}_{0}  \ , 
\end{align}
having made the standard phase choice that $\Delta^{(1)}_l(\cdot,K)$ is orthogonal to  $\Phi_l^{(0)}(\cdot,K)$.

The Hamiltonian matrix elements can be written, for $(n,n') \in (\mathbb{N}^{\ast})^2$, 
\begin{align} \label{eq:matrix_elements}
\langle \lambda',n' | \mathcal{H}^{(1)} | \lambda,n \rangle  = \ \ \
& \frac{iB}{2} \ ( \alpha^{\ast}_{\lambda',n'} \alpha_{\lambda,n}  - \beta^{\ast}_{\lambda',n'} \beta_{\lambda,n} ) \ \delta_{n',n}  \\
 + \ & iB \ ( \alpha^{\ast}_{\lambda',n'} \alpha_{\lambda,n} \ I_{n'-1,n-1} - \beta^{\ast}_{\lambda',n'} \beta_{\lambda,n} \ I_{n',n}  ) \nonumber \\
 + \ & m \ ( \beta^{\ast}_{\lambda',n'} \alpha_{\lambda,n} \ J_{n',n-1} - \alpha^{\ast}_{\lambda',n'} \beta_{\lambda,n} \ J_{n'-1,n} ) \ , \nonumber
\end{align}
\noindent
where $\delta_{n',n}$ is the Kronecker symbol and where, for $(n,n') \in \mathbb{N}^2$,
\begin{align}
\label{I} I_{n',n} & = \int_{-\infty}^{+\infty} dX \ X \ \phi^{HO}_{n'}(X) \ \partial_X [\phi^{HO}_{n}(X)] \ , \\
\label{J} J_{n',n} & = \int_{-\infty}^{+\infty} dX \  \phi^{HO}_{n'}(X) \ \partial_X [ \phi^{HO}_{n}(X)]  \ .
\end{align}
\noindent
Expression (\ref{eq:matrix_elements}) can be extended to $l=0$ and $l'=0$, having preliminarily defined
\begin{equation} \label{eq:alpha_0_beta_0}
 \alpha_0 = 0 \ , \ \ \  \ \ \ \ \  \beta_0 = 1 \ .
\end{equation}

Thanks to the following recurrence relations,
\begin{align}
a \sqrt{2}  \ \partial_X [\phi^{HO}_n(X)] & = \sqrt{n} \ \phi^{HO}_{n-1}(X) - \sqrt{n+1} \ \phi^{HO}_{n+1}(X) \ , \\
2 X H_n(X) & = 2n \ H_{n-1}(X) + H_{n+1}(X) \ ,
\end{align}
we can compute integrals (\ref{I}) and (\ref{J}), and finally find the following expressions for the Hamiltonian matrix elements:

\begin{align}
\langle \lambda',n' | \mathcal{H}^{(1)} | \lambda,n \rangle =
\left\{
  \begin{array}{l l}
  \mbox{sgn}(B) \sqrt{n-1} \left[ \frac{iB}{2} (\alpha^{\ast}_{\lambda',n'} \alpha_{\lambda,n} \sqrt{n'} - \beta^{\ast}_{\lambda',n'} \beta_{\lambda,n} \sqrt{n}) + \frac{m}{a\sqrt{2}} \beta^{\ast}_{\lambda',n'} \alpha_{\lambda,n} \right] &
  \mathrm{for} \ n' = n-2 \\
  - \mbox{sgn}(B) \frac{m \sqrt{n}}{a \sqrt{2}}(\alpha^{\ast}_{\lambda',n} \beta_{\lambda,n}  + \beta^{\ast}_{\lambda',n} \alpha_{\lambda,n} ) &
  \mathrm{for} \ n' = n \\
    \mbox{sgn}(B) \sqrt{n+1} \left[ \frac{iB}{2} (-\alpha^{\ast}_{\lambda',n'} \alpha_{\lambda,n} \sqrt{n} + \beta^{\ast}_{\lambda',n'} \beta_{\lambda,n} \sqrt{n'}) + \frac{m}{a\sqrt{2}} \alpha^{\ast}_{\lambda',n'} \beta_{\lambda,n}  \right] &
  \mathrm{for} \ n' = n+2 \\
  0 & \mathrm{otherwise} \ ,
  \end{array}
\right. 
\end{align}
which again can be extended to $l=0$ and $l'=0$.
Hence, the sums in Eqs. (\ref{eigenstate1_0}) and (\ref{eigenstate1}) reduce to sums over $n' = n \pm 2$, and, because of the relation between $\alpha_l$ and $\beta_l$ (Eqs. (\ref{eq:alpha_beta}) and \ref{eq:alpha_0_beta_0})), the first-order corrections to the energies vanishe, {\sl i.e.}
$\langle l | \mathcal{H}^{(1)} | l \rangle = 0$ .


\end{document}